\documentclass[11pt, letterpaper, onecolumn, peerreview]{IEEEtran}
\usepackage{exscale}
\usepackage{color}
\usepackage{colordvi}
\usepackage{graphicx}
\usepackage{amsmath}
\usepackage{amssymb}
\usepackage{epsfig}
\usepackage{epic}

\newtheorem{theorem}{Theorem}[section]
\newtheorem{lemma}{Lemma}[section]

\newtheorem{definition}{Definition}[section]

\begin{document}
\title{Balancing forward and feedback error correction for erasure
  channels with unreliable feedback} 

\author{Anant Sahai\footnote{Wireless Foundations, Department of
    Electrical Engineering and Computer Science at the University of
    California at Berkeley. A preliminary version of these results was
    presented at the 2007 ITA Workshop in San Diego.}  \\ {\small
    sahai@eecs.berkeley.edu}}

\maketitle

\begin{abstract} 
  The traditional information theoretic approach to studying feedback
  is to consider ideal instantaneous high-rate feedback of the channel
  outputs to the encoder. This was acceptable in classical work
  because the results were negative: Shannon pointed out that even
  perfect feedback often does not improve capacity and in the context
  of symmetric DMCs, Dobrushin showed that it does not improve the
  fixed block-coding error exponents in the interesting high
  rate regime. However, it has recently been shown that perfect
  feedback {\bf does} allow great improvements in the asymptotic
  tradeoff between end-to-end delay and probability of error, even for
  symmetric channels at high rate. Since gains are claimed with ideal
  instantaneous feedback, it is natural to wonder whether these
  improvements remain if the feedback is unreliable or otherwise
  limited.

  Here, packet-erasure channels are considered on both the forward and
  feedback links. First, the feedback channel is considered as a given
  and a strategy is given to balance forward and feedback error
  correction in the suitable information-theoretic limit of long
  end-to-end delays. At high enough rates, perfect-feedback
  performance is asymptotically attainable despite having only
  unreliable feedback! Second, the results are interpreted in the
  zero-sum case of ``half-duplex'' nodes where the allocation of
  bandwidth or time to the feedback channel comes at the direct
  expense of the forward channel. It turns out that even here,
  feedback is worthwhile since dramatically lower asymptotic delays
  are possible by appropriately balancing forward and feedback error
  correction.

  The results easily generalize to channels with strictly positive
  zero-undeclared-error capacities. 
\end{abstract}


\IEEEpeerreviewmaketitle

\section{Introduction}
The three most fundamental parameters when it comes to reliable data
transport are probability of error, end-to-end system delay, and data
rate. Error probability is critical because a low probability of
bit error lies at the heart of the digital revolution justified by the
source/channel separation theorem. Delay is important because it is
the most basic cost that a system must pay in exchange for reliability
--- it allows the laws of large numbers to be harnessed to smooth out
the variability introduced by random communication channels. 

Data rate, while apparently the most self-evidently important of the
three, also brings out the seemingly pedantic question of units: do we
mean bits per second or bits per channel use? In a point-to-point
setting without feedback, there is an unambiguous mapping between the
two of them given by the number of channel uses per second. When
feedback is allowed, an ambiguity arises: how should the feedback
channel uses be accounted for? Are they ``free'' or do they need to be
counted?

Traditionally, the approach has been to consider feedback as ``free''
because the classical results showed that in many cases even free
feedback does not improve the capacity \cite{ShannonZeroError}, and in
the fixed-length block-coding context, does not even improve the
tradeoff between the probability of error and delay in the high-rate
regime for symmetric channels \cite{DobrushinReliability,
  BerlekampThesis}. (See \cite{OurUpperBoundPaper} for a detailed
review of this literature.) Thus, the answer was simple: if feedback
comes at any cost, it is not worth using for memoryless channels.

In \cite{OurUpperBoundPaper}, we recently showed that perfect feedback
is indeed quite useful in a ``streaming'' context if we are willing to
use non-block codes to implement our communication system. In the
natural setting of a message stream being produced at a fixed
(deterministic) rate of $R$ bits per second, feedback {\em does}
provide\footnote{This overturned Pinsker's incorrect assertion in
  Theorem 8 of \cite{PinskerNoFeedback} that feedback gives no
  asymptotic advantage in this nonblock setting.} a tremendous
advantage in terms of the tradeoff between end-to-end delay and the
probability of bit error. As the desired probability of error tends to
zero, feedback reduces the end-to-end delay by a {\em factor} that
approaches infinity as the desired rate approaches capacity. In
\cite{OurUpperBoundPaper}, the resulting fixed-delay error exponents
with feedback are calculated exactly for erasure channels and channels
with strictly positive zero-error capacity. For general DMCs,
\cite{OurUpperBoundPaper} gives a general upper bound (the
uncertainty-focusing bound) along with a suboptimal construction that
substantially outperforms codes without feedback at high rates.

Once it is known that perfect feedback is very useful, it is natural
to ask whether this advantage persists if feedback is costly,
rate-limited, or unreliable in some way. After all, the real question
is not whether perfect feedback would be useful but how imperfect
feedback is worth designing in real communication systems. This has
long been recognized as the Achilles' Heel for the
information-theoretic study of feedback. Bob Lucky in
\cite{LuckySurvey73} stated it dramatically:
\begin{quotation}
{\small  Feedback communications was an area of intense activity in
  1968.\ldots A number of authors had shown constructive, even
  simple, schemes using noiseless feedback to achieve Shannon-like
  behavior\ldots The situation in 1973 is dramatically
  different.\ldots {\em The subject itself seems to be a
  burned out case.} \ldots

  In extending the simple noiseless feedback model to allow for more
  realistic situations, such as noisy feedback channels,
  bandlimited channels, and peak power constraints, theorists
  discovered a certain ``brittleness'' or sensitivity in their
  previous results.}
\end{quotation}

The literature on imperfect feedback in the context of memoryless
channels is relatively thin. Schulman and others have studied
interaction over noisy channels in the context of distributed
computation \cite{Schulman, Rajagopalan}. The computational agents can
only communicate with each other through noisy channels in both
directions. In Schulman's formulation, neither capacity nor end-to-end
delay is a question of major interest since constant factor slowdowns
are seemingly unavoidable. The focus is instead on making sure that
the computation succeeds and that the slowdown factor does not scale
with problem size (as it would for a purely repetition based
strategy).

On the reliability side before recently, all the limited successes
were for continuous-alphabet AWGN type-channels following the
Schalkwijk/Kailath model from \cite{Kailath66, Schalkwijk66}. Kashyap
in \cite{KashyapNoisyFeedback} introduced a scheme that tolerated
noise on the feedback link, but it used asymptotically infinite (in
the block-length) power on the feedback link to overcome it. It is
only the work by Kramer \cite{IntermittentFeedback} and Lavenburg
\cite{LavenburgOrthogonal, LavenburgRepetitive} that worked with
finite average power in both directions. But these were all cases in
which the average nature of the power-constraint played an important
role. Recently, the AWGN story with noisy feedback has also attracted
the interest of Weissman, Lapidoth and Kim in
\cite{WeissmanKailathLecture}, who rigorously proved a strongly
negative folk result for the case of {\em uncoded} channel-output
feedback corrupted by arbitrarily low levels of Gaussian noise. At the
same time, \cite{MartinsKailathLecture} showed that if the feedback
noise has bounded support, then techniques similar to those of
\cite{ControlPartI} could preserve reliability gains, but only at the
price of having to back away from the capacity of the forward link.

For finite alphabets, we have recently had some success in showing
robustness to imperfect feedback in the ``soft deadlines'' context
where the decoder is implicitly allowed to postpone making a decision,
as long as it does not do so too often. With perfect feedback, this
has traditionally been considered in the variable-block-length setting
where Burnashev's bound of \cite{burnashev} gives the ultimate limit
with perfect feedback and Yamamoto-Itoh's scheme of \cite{yamamoto}
provides the baseline architecture. \cite{ITWpaper} showed that if the
feedback channel was noisy, but of very high quality, then the loss
relative to the Burnashev bound could be quite small by appropriately
using anytime codes and pipelining. \cite{DraperAllerton05} allowed
bursty noiseless feedback, but constrained its overall rate to show
that by using hashing ideas, something less than full channel output
feedback could be used while still attaining the Burnashev bound. The
ideas of \cite{ITWpaper, DraperAllerton05} were combined in
\cite{DraperISIT06Paper} to show that it was possible to get
reliability within a factor of two of the Burnashev bound as long as
the noisy feedback channel's capacity was higher than the capacity of
the forward link and this was further tightened in
\cite{DraperECCPaper} to a factor that approaches one as the target
rate approached capacity. This story culminates in
\cite{DraperAllerton06} where it is shown that from the system-level
perspective, Burnashev's bound is not the relevant target. Instead,
Kudryashov's performance with noiseless feedback in
\cite{KudryashovPPI} (better than the Burnashev Bound) can in fact be
asymptotically attained robustly as long as the feedback channel's
capacity is larger than the target reliability.


The focus here is on the problem of {\em fixed} rate and {\em fixed}
end-to-end delay in the style of \cite{OurUpperBoundPaper} where the
decoder is not allowed to postpone making a decision. This paper
restricts attention to the case of memoryless packet-erasure channels
where the feedback path is also an unreliable packet-erasure
channel. Recently, Massey \cite{MasseyLecture} has had some
interesting thoughts on this problem, but he claims no asymptotic
reliability gains for uncertain feedback over no feedback. The issue
of balancing forward and feedback error correction has also attracted
attention in the networking community (see eg~\cite{OverQoS}), but the
focus there is not purely on reliability or delay but is mixed with
the issue of adapting to bursty channel variation as well as fairness
with other streams.

Section~\ref{sec:definitions} establishes the notation and states the
main results, with the proofs following in subsequent sections. The
results are stated for the concrete case of packet-erasure
channels. Section~\ref{sec:definitions} also plots the performance for
some examples and compares the results with the baseline approaches of
only using forward error correction and just using feedback for
requesting retransmissions at the individual packet level.

In Section~\ref{sec:freefeedback}, the feedback channel uses are
considered ``free'' in that they do not compete with forward channel
uses for access to the underlying communication medium. Adapting
arguments from \cite{OurUpperBoundPaper}, it is shown that
asymptotically, perfect feedback performance can be attained even with
unreliable feedback. Because the uncertainty-focusing bound of
\cite{OurUpperBoundPaper} is met at rates that are high enough, it is
known that this is essentially optimal. If the target rate is too low,
then the dominant error event for the scheme turns out to be the
feedback channel going into complete outage and erasing every packet.

At the end of Section~\ref{sec:freefeedback}, it is noted that the
main result generalizes naturally from packet-erasure channels to DMCs
whose zero-undetected-error capacity is equal to their Shannon
capacity. As a bonus\footnote{Dear Reviewers: the result referred to
  here does not really fit in with the overall theme of unreliable
  feedback, but is placed in this paper since the techniques used are
  common.}, the same techniques give rise to a generalization of
Theorem~3.4 of \cite{OurUpperBoundPaper} and show the achievability
with perfect feedback of the symmetric uncertainty-focusing bound at
high rate for any channel whose probability matrix contains a
nontrivial zero.

Section~\ref{sec:costlyfeedback} reinterprets the results of the
previous section to address the question of ``costly'' feedback in
that both the data rate and delay are measured not relative to forward
channel uses, but relative to the sum of feedback and forward channel
uses. This models the case when the feedback channel uses the same
underlying communication resource as the forward channel in a zero-sum
way (e.g. time-division or frequency-division in wireless
networks). It is shown that there is a tremendous advantage to using
some of the channel uses for feedback.

\section{Definitions and main results}
\label{sec:definitions}

\begin{figure}
\begin{center}
\setlength{\unitlength}{3000sp}%
\begingroup\makeatletter\ifx\SetFigFont\undefined%
\gdef\SetFigFont#1#2#3#4#5{%
  \reset@font\fontsize{#1}{#2pt}%
  \fontfamily{#3}\fontseries{#4}\fontshape{#5}%
  \selectfont}%
\fi\endgroup%
\begin{picture}(7155,1674)(436,-2623)
\thinlines
{\color[rgb]{0,0,0}\put(1051,-1561){\framebox(900,600){}}
}%
\put(1501,-1336){\makebox(0,0)[b]{\smash{{\SetFigFont{12}{14.4}{\familydefault}{\mddefault}{\updefault}{\color[rgb]{0,0,0}${\cal E}_f$}%
}}}}
{\color[rgb]{0,0,0}\put(2851,-1561){\framebox(1500,600){}}
}%
\put(3601,-1261){\makebox(0,0)[b]{\smash{{\SetFigFont{9}{6}{\familydefault}{\mddefault}{\updefault}{\color[rgb]{0,0,0}Forward Channel}%
}}}}
\put(3601,-1456){\makebox(0,0)[b]{\smash{{\SetFigFont{12}{14.4}{\familydefault}{\mddefault}{\updefault}{\color[rgb]{0,0,0}$P_f$}%
}}}}
{\color[rgb]{0,0,0}\put(601,-1261){\vector( 1, 0){450}}
}%
{\color[rgb]{0,0,0}\put(1951,-1261){\vector( 1, 0){900}}
}%
{\color[rgb]{0,0,0}\put(4351,-1261){\vector( 1, 0){1800}}
}%
{\color[rgb]{0,0,0}\put(5401,-1261){\vector( 0,-1){750}}
}%
{\color[rgb]{0,0,0}\put(4951,-2311){\vector(-1, 0){600}}
}%
{\color[rgb]{0,0,0}\put(2851,-2311){\line(-1, 0){1350}}
\put(1501,-2311){\vector( 0, 1){750}}
}%
{\color[rgb]{0,0,0}\put(7051,-1261){\vector( 1, 0){450}}
}%
{\color[rgb]{0,0,0}\put(2851,-2611){\framebox(1500,600){}}
}%
{\color[rgb]{0,0,0}\put(4951,-2611){\framebox(900,600){}}
}%
{\color[rgb]{0,0,0}\put(6151,-1561){\framebox(900,600){}}
}%
\put(6601,-1336){\makebox(0,0)[b]{\smash{{\SetFigFont{12}{14.4}{\familydefault}{\mddefault}{\updefault}{\color[rgb]{0,0,0}${\cal D}$}%
}}}}
\put(5401,-2386){\makebox(0,0)[b]{\smash{{\SetFigFont{12}{14.4}{\familydefault}{\mddefault}{\updefault}{\color[rgb]{0,0,0}${\cal E}_b$}%
}}}}
\put(3601,-2311){\makebox(0,0)[b]{\smash{{\SetFigFont{9}{6}{\familydefault}{\mddefault}{\updefault}{\color[rgb]{0,0,0}Feedback Channel}%
}}}}
\put(3601,-2506){\makebox(0,0)[b]{\smash{{\SetFigFont{12}{14.4}{\familydefault}{\mddefault}{\updefault}{\color[rgb]{0,0,0}$P_b$}%
}}}}
\put(2401,-1186){\makebox(0,0)[b]{\smash{{\SetFigFont{12}{14.4}{\familydefault}{\mddefault}{\updefault}{\color[rgb]{0,0,0}$X$}%
}}}}
\put(5401,-1186){\makebox(0,0)[b]{\smash{{\SetFigFont{12}{14.4}{\familydefault}{\mddefault}{\updefault}{\color[rgb]{0,0,0}$Y$}%
}}}}
\put(451,-1336){\makebox(0,0)[b]{\smash{{\SetFigFont{12}{14.4}{\familydefault}{\mddefault}{\updefault}{\color[rgb]{0,0,0}$B$}%
}}}}
\put(7576,-1336){\makebox(0,0)[b]{\smash{{\SetFigFont{12}{14.4}{\familydefault}{\mddefault}{\updefault}{\color[rgb]{0,0,0}$\widehat{B}$}%
}}}}
\end{picture}%
\caption{The problem of communication with unreliable feedback. For
  the most part in this paper, both of the unreliable channels $P_f$
  and $P_b$ are packet erasure channels that are independent of each
  other.}
\label{fig:feedbackblockdiagram}
\end{center}
\end{figure}
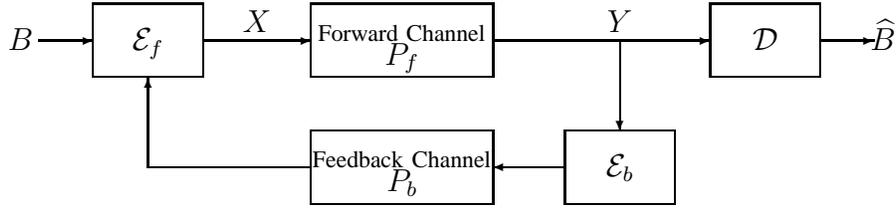

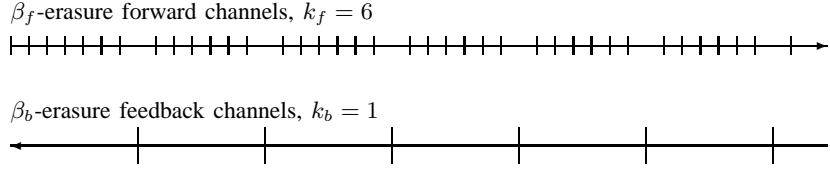
\begin{figure}
\begin{center}
\setlength{\unitlength}{3000sp}%
\begingroup\makeatletter\ifx\SetFigFont\undefined%
\gdef\SetFigFont#1#2#3#4#5{%
  \reset@font\fontsize{#1}{#2pt}%
  \fontfamily{#3}\fontseries{#4}\fontshape{#5}%
  \selectfont}%
\fi\endgroup%
\begin{picture}(6774,1627)(-11,-1469)
\thinlines
{\color[rgb]{0,0,0}\put(  1,-136){\line( 0,-1){150}}
}%
{\color[rgb]{0,0,0}\put(151,-136){\line( 0,-1){150}}
}%
{\color[rgb]{0,0,0}\put(301,-136){\line( 0,-1){150}}
}%
{\color[rgb]{0,0,0}\put(451,-136){\line( 0,-1){150}}
}%
{\color[rgb]{0,0,0}\put(601,-136){\line( 0,-1){150}}
}%
{\color[rgb]{0,0,0}\put(751,-136){\line( 0,-1){150}}
}%
{\color[rgb]{0,0,0}\put(901,-136){\line( 0,-1){150}}
}%
{\color[rgb]{0,0,0}\put(1201,-136){\line( 0,-1){150}}
}%
{\color[rgb]{0,0,0}\put(1351,-136){\line( 0,-1){150}}
}%
{\color[rgb]{0,0,0}\put(1501,-136){\line( 0,-1){150}}
}%
{\color[rgb]{0,0,0}\put(1651,-136){\line( 0,-1){150}}
}%
{\color[rgb]{0,0,0}\put(1801,-136){\line( 0,-1){150}}
}%
{\color[rgb]{0,0,0}\put(1951,-136){\line( 0,-1){150}}
}%
{\color[rgb]{0,0,0}\put(2251,-136){\line( 0,-1){150}}
}%
{\color[rgb]{0,0,0}\put(2401,-136){\line( 0,-1){150}}
}%
{\color[rgb]{0,0,0}\put(2551,-136){\line( 0,-1){150}}
}%
{\color[rgb]{0,0,0}\put(2701,-136){\line( 0,-1){150}}
}%
{\color[rgb]{0,0,0}\put(2851,-136){\line( 0,-1){150}}
}%
{\color[rgb]{0,0,0}\put(3001,-136){\line( 0,-1){150}}
}%
{\color[rgb]{0,0,0}\put(3301,-136){\line( 0,-1){150}}
}%
{\color[rgb]{0,0,0}\put(3451,-136){\line( 0,-1){150}}
}%
{\color[rgb]{0,0,0}\put(3601,-136){\line( 0,-1){150}}
}%
{\color[rgb]{0,0,0}\put(3751,-136){\line( 0,-1){150}}
}%
{\color[rgb]{0,0,0}\put(3901,-136){\line( 0,-1){150}}
}%
{\color[rgb]{0,0,0}\put(4051,-136){\line( 0,-1){150}}
}%
{\color[rgb]{0,0,0}\put(4351,-136){\line( 0,-1){150}}
}%
{\color[rgb]{0,0,0}\put(4501,-136){\line( 0,-1){150}}
}%
{\color[rgb]{0,0,0}\put(4651,-136){\line( 0,-1){150}}
}%
{\color[rgb]{0,0,0}\put(4801,-136){\line( 0,-1){150}}
}%
{\color[rgb]{0,0,0}\put(4951,-136){\line( 0,-1){150}}
}%
{\color[rgb]{0,0,0}\put(5101,-136){\line( 0,-1){150}}
}%
{\color[rgb]{0,0,0}\put(5401,-136){\line( 0,-1){150}}
}%
{\color[rgb]{0,0,0}\put(5551,-136){\line( 0,-1){150}}
}%
{\color[rgb]{0,0,0}\put(5701,-136){\line( 0,-1){150}}
}%
{\color[rgb]{0,0,0}\put(5851,-136){\line( 0,-1){150}}
}%
{\color[rgb]{0,0,0}\put(6001,-136){\line( 0,-1){150}}
}%
{\color[rgb]{0,0,0}\put(6151,-136){\line( 0,-1){150}}
}%
{\color[rgb]{0,0,0}\put(6451,-136){\line( 0,-1){150}}
}%
{\color[rgb]{0,0,0}\put(6751,-1036){\vector( -1, 0){6750}}
}%
{\color[rgb]{0,0,0}\put(1051,-886){\line( 0,-1){300}}
}%
{\color[rgb]{0,0,0}\put(2101,-886){\line( 0,-1){300}}
}%
{\color[rgb]{0,0,0}\put(3151,-886){\line( 0,-1){300}}
}%
{\color[rgb]{0,0,0}\put(4201,-886){\line( 0,-1){300}}
}%
{\color[rgb]{0,0,0}\put(5251,-886){\line( 0,-1){300}}
}%
{\color[rgb]{0,0,0}\put(6301,-886){\line( 0,-1){300}}
}%
{\color[rgb]{0,0,0}\put(  1,-211){\vector( 1, 0){6750}}
}%
\put(  1, 14){\makebox(0,0)[lb]{\smash{\SetFigFont{9}{6}{\rmdefault}{\mddefault}{\updefault}{\color[rgb]{0,0,0}$\beta_f$-erasure forward channels, $k_f=6$ }%
}}}
\put(
1,-811){\makebox(0,0)[lb]{\smash{\SetFigFont{9}{6}{\rmdefault}{\mddefault}{\updefault}{\color[rgb]{0,0,0} $\beta_b$-erasure feedback channels, $k_b=1$ }%
}}}
\end{picture}
\caption{Feedback and forward channel uses. The forward channel has
  packet-erasure probability $\beta_f$ while the feedback channel has
  packet-erasure probability $\beta_b$. In this picture $k_f=6,k_b=1$
  So a fraction $\frac{1}{7}$ of the total channel uses are allocated
  to feedback. } 
\label{fig:feedbacktimeline}
\end{center}
\end{figure}

The problem and notation is illustrated in Figures
\ref{fig:feedbackblockdiagram} and
\ref{fig:feedbacktimeline}. Formally: 

\begin{definition} \label{def:packeterasurechannel}
A {\em $C_p$-bit $\beta$-erasure channel} refers to a
discrete memoryless channel that accepts packets $X(t)$ consisting of
$C_p$ bits (thought of as integers from $0$ to $2^{C_p}-1$ or strings
from $\{0,1\}^{C_p}$) per packet
as inputs and either delivers the entire packet perfectly $Y(t) = X(t)$
with probability $1-\beta$ or erases the whole packet $Y(t) = -1$ 
with probability $\beta$. 
\end{definition}
\vspace{0.1in}
\begin{definition} \label{def:ratiochannel}
The {\em $(k_f,k_b,C_f,C_b,\beta_f,\beta_b)$ problem} consists of a
system in which one cycle of interaction between encoders and decoders
consists of $k_f$ independent packets being sent along a $C_f$-bit
$\beta_f$-erasure channel along the forward direction and $k_b$
independent packets being sent along a $C_b$-bit $\beta_b$-erasure
channel along the reverse direction. 

A {\em feedback encoder ${\cal E}_b$} for this problem is a
sequence of maps ${\cal E}_{b,t}$. The range of each map is $k_b$ packets
$(X_b(k_b t + 1), X_b(k_b t + 2), \ldots, X_b(k_b t + k_b))$ consisting
of $C_b$ bits each. The $t$-th map takes as input all the available
forward channel outputs $(Y_f(1), \ldots, Y_f(k_f t))$ so far.

A {\em rate $R$ forward encoder ${\cal E}_f$} for this problem is also a
sequence of maps ${\cal E}_{f,t}$. The range of each map is $k_f$ packets
$(X_f(k_f t + 1), X_f(k_f t + 2), \ldots, X_f(k_f t + k_f))$ consisting
of $C_f$ bits each. The $t$-th map takes as input all the available
feedback channel outputs $(Y_b(1), \ldots, Y_b(k_b t))$ as well as the
message bits $B_1^{\lfloor (R k_f C_f) t \rfloor}$ so far.

The rate $R$ above is in terms of forward channel uses only and is
normalized in units of $C_f$-bit packets. The rate $R'$ in terms of
overall channel uses is $R' = R \frac{k_f}{k_f + k_b}$. The rate $\bar{R}$
in terms of weighted channel uses is $\bar{R} = R \frac{k_f C_f}{k_f C_f +
  k_b C_b}$.

A {\em delay $d$ rate $R$ decoder} is a sequence of maps ${\cal
  D}_i$. The range of each map is an estimate $\widehat{B}_i$ for the
$i$-th bit taken from $\{0,1\}$. The $i$-th map takes as input the
available channel outputs $(Y_f(1), Y_f(2), \ldots, Y_f(\lceil
\frac{i}{R k_f C_f} \rceil + d k_f))$. This means that it can see $d
k_f$ channel uses beyond when the bit to be estimated first had the
potential to influence the channel inputs.

Just as rate can be expressed in different units, so can delay. The
delay $d$ above is in terms of forward channel uses only. The delay
$d'$ in terms of overall channel uses is $d' = d \frac{k_f +
  k_b}{k_f}$. The delay $\bar{d}$ in terms of weighted channel uses is
$\bar{d} = d \frac{k_f C_f + k_b C_b}{k_f C_f}$.

All encoders and decoders are assumed to be randomized and have access
to infinite amounts of common randomness that is independent of both
the messages as well as the channel unreliability.
\end{definition}
\vspace{0.1in}

The notation above captures the real flexibility of interest. The
distinction between $C_f$ and $C_b$ allows for forward and feedback
packets to be of different size. The distinction between $k_f$ and
$k_b$ summarizes the relative width of the forward and feedback
pipes. Similarly, the distinction between $\beta_f$ and $\beta_b$
allows for the channel to be more or less reliable in the
different directions.  

The three kinds of units correspond to three different ways of
thinking about the problem.  
\begin{itemize}
 \item The unadorned $R$ and $d$ take the traditional approach of
       considering the feedback to be entirely free, although
       there is now a limited amount of it and it is imperfect.

 \item The $R'$ and $d'$ consider feedback to be costly, but in terms
       of channel uses only. The relative size of the packets is
       considered unimportant. It is clear that as long as some
       feedback is present, $R' < R$ and $d' > d$. These metrics give
       an incentive to use less feedback if possible.

 \item The $\bar{R}$ and $\bar{d}$ also consider the relative size of the
       packets significant. These metrics give an incentive to use
       shorter feedback packets.
\end{itemize}

It is even interesting to consider combinations of metrics. The
combination of $\bar{R}$ and $d'$ is particularly interesting since it
corresponds to the case when feedback is implemented as bits stolen
from message-carrying packets coming in the reverse direction. Using
$\bar{R}$ gives an incentive to make the number of bits taken small, but
using $d'$ captures the fact that the delay in real time units
includes the full lengths of both the intervening forward and feedback
packets. 

\begin{definition} \label{def:achievable} The fixed-delay error
  exponent $\alpha$ is asymptotically {\em achievable} at message rate
  $R$ across a noisy channel if for every delay $d_j$ in some
  increasing sequence $d_j \rightarrow \infty$ there exist rate $R$
  encoders $({\cal E}_f^{j}, {\cal E}_b^{j})$ and delay $d_j$ decoders
  ${\cal D}^{j}$ that satisfy the following properties when used with
  input bits $B_i$ drawn from iid fair coin tosses.
\begin{enumerate} 
\item For the $j$-th code, there exists an $\epsilon_j < 1$ so that
  ${\cal P}(B_i \neq \widehat{B}_i(d_j)) \leq \epsilon_j$ for every
  bit position $i \geq 1$. The $\widehat{B}_i(d_j)$ represents the
  delay $d_j$ estimate of $B_i$ produced by the $({\cal E}_f^{j},
  {\cal E}_b^{j}, {\cal D}^j)$ triple connected through the channels in
  question.

 \item $\lim_{j \rightarrow \infty} \frac{-\ln \epsilon_j}{d_j k_f} \geq
       \alpha$.
\end{enumerate}

The exponent $\alpha$ is asymptotically {\em achievable
  universally over delay} or in an {\em anytime fashion} if a single
encoder pair $({\cal E}_f^{j}, {\cal E}_b^{j})$ can be used
simultaneously for all sufficiently long $d_j$. Only the decoders
${\cal D}^j$ have to change with the target delay $d_j$.

The error exponents $\alpha'$ and $\bar{\alpha}$ are defined analogously
but use the $d_j'$ and $\bar{d}_j$ versions of delay.
\end{definition}
\vspace{0.1in}

\subsection{Main Results}
\begin{theorem}  \label{thm:nolistbasictheorem}
Given the $(k_f,k_b,C_f,C_b,\beta_f,\beta_b)$ problem with
$k_f,k_b \geq 1$, forward packet size $C_f \geq 1$, and feedback
packet size $C_b \geq 1$, it is possible to asymptotically achieve all
fixed-delay reliabilities
\begin{equation} \label{eqn:basicboundforalphalist1}
\alpha < \min\left(-\frac{k_b}{k_f} \ln \beta_b, E_0(C_f, 1)\right)
\end{equation}
where the Gallager function for the forward channel is
\begin{equation} \label{eqn:Enoughtdefinition}
E_0(C_f, \rho) = -\ln(\beta_f + 2^{-\rho C_f}(1-\beta_f)),
\end{equation}
as long as the rate $R$ in normalized $C_f$ units satisfies
\begin{equation} \label{eqn:focusingbound}
R < \frac{\alpha}{\alpha + \ln \left(\frac{1-\beta_f}{1-\exp(\alpha)
      \beta_f}\right)}. 
\end{equation}
Furthermore, these fixed-delay reliabilities are obtained in an anytime fashion. 
\end{theorem}
{\em Proof:} See Section~\ref{sec:nolistproof}.
\vspace{0.1in}

The uncertainty-focusing bound from
\cite{OurUpperBoundPaper} for this problem assuming {\em perfect
  feedback} is easily calculated to be given by
(\ref{eqn:focusingbound}) but it holds for all $0 < \alpha < -\ln
\beta_f$. Since lower reliabilities are associated with higher rates,
this shows that the result of Theorem~\ref{thm:nolistbasictheorem} is
asymptotically optimal at high enough rates. The feedback packet size
needs to be only one bit and there is similarly no
restriction on the size of forward packets. Since $\lim_{C_f
  \rightarrow \infty} E_0(C_f, 1) = -\ln \beta_f$, the sense of high
enough rates given by (\ref{eqn:basicboundforalphalist1}) depends only
on the relative frequency and reliability of the feedback link as the
forward packet size tends to infinity. 

When the packet sizes are at least two bits long in both directions,
asymptotic delay performance can be slightly improved at low rates.

\begin{theorem}  \label{thm:listbasictheorem}
Given the $(k_f,k_b,C_f,C_b,\beta_f,\beta_b)$ problem with
$k_f,k_b \geq 1$, $C_f \geq 2$, and feedback packet size $C_b \geq 2$, it
is possible to asymptotically achieve all fixed-delay reliabilities
\begin{equation} \label{eqn:basicboundforalphalistarbitrary}
\alpha < -\frac{k_b}{k_f} \ln \beta_b
\end{equation}
as long as the rate $R$ in normalized $C_f$ units satisfies
\begin{equation} \label{eqn:basicboundforalphalistbig}
R < \left(\frac{C_f - 1}{C_f}\right)
\frac{\alpha}{\alpha + \ln \left(\frac{1-\beta_f}{1-\exp(\alpha) \beta_f}\right)}.
\end{equation}
Furthermore, these fixed-delay reliabilities are obtained in an anytime
fashion.
\end{theorem}
{\em Proof:} See Section~\ref{sec:listproof}.
\vspace{0.1in}

The upper-limit on reliability for the scheme given by
(\ref{eqn:basicboundforalphalistarbitrary}) corresponds to the event
that the feedback channel erases every feedback packet for the entire
duration of $d$ cycles. If $k_f$ and $k_b$ could be chosen by the
system designer, this constraint could be made non-binding simply by
choosing $\frac{k_b}{k_f}$ large enough. However, if there is such
flexibility, it is only fair to also penalize based on the total
resources used, rather than only penalizing forward channel uses.

\begin{theorem}  \label{thm:balancedtheorem}
Given only $(C_f \geq 1, C_b \geq 1,\beta_f > 0, \beta_b > 0)$, the
$k_f > 0$ and $k_b > 0$ can be chosen to asymptotically achieve all
$(R',\alpha')$ pairs that are contained within the parametric region: 
\begin{eqnarray}
\alpha' & < & E_0'(C_f,\rho), \nonumber \\
R'      & < & \frac{E_0'(C_f, \rho)}{\rho C_f \ln
  2} \label{eqn:balancedRegionNoList}
\end{eqnarray}
where 
\begin{equation} \label{eqn:balancedEnought}
E_0'(C_f, \rho) = \left((-\ln \beta_f)^{-1} + (E_0(C_f,\rho))^{-1}
\right)^{-1} 
\end{equation}
and the Gallager function $E_0(C_f,\rho)$ is defined in
(\ref{eqn:Enoughtdefinition}) and $\rho$ ranges from $0$ to $1$. 

If furthermore $C_f \geq 2, C_b \geq 2$, then the following
region is also attainable:
\begin{eqnarray}
\alpha' & < & E_0'(C_f-1,\rho), \nonumber \\
R'      & < & \frac{E_0'(C_f - 1, \rho)}{\rho C_f \ln 2} \label{eqn:balancedRegionList} 
\end{eqnarray}
with $\rho$ ranging from $0$ to $\infty$. 

If $C_b$ stays constant while $C_f$ can be chosen as large as desired,
then $(\bar{R}, \alpha')$ in
\begin{eqnarray}
\alpha' & < & E_0'(C_f, \rho), \nonumber \\
\bar{R}     & < & \frac{E_0(C_f,\rho)}{\rho C_f \ln 2} \label{eqn:balancedRegionMixedTradeoff}
\end{eqnarray}
can be achieved where $\rho \in [0,1]$. If the $(\bar{R},\bar{\alpha})$
tradeoff is desired, use (\ref{eqn:balancedRegionMixedTradeoff}) for
$\bar{R}$ with  $\bar{\alpha} < E_0(C_f, \rho)$.

All of these fixed-delay reliabilities are obtained in an anytime fashion. 
\end{theorem}
{\em Proof:} See Section~\ref{sec:costlyfeedback}.
\vspace{0.1in}


\subsection{Pure strategies and comparison plots}

To understand the implications of Theorems
\ref{thm:nolistbasictheorem}, \ref{thm:listbasictheorem} and
\ref{thm:balancedtheorem}, it is useful to compare them to what
would be obtained using strategies for reliable communication that use
only feedback error correction or only forward error correction. 

\subsubsection{Pure forward error correction}
The simplest approach is to ignore the feedback and just use a
fixed-delay code. \cite{JelinekSequential, ForneyML} reveal that
infinite-constraint-length convolutional codes achieve the
random-coding error exponent $E_r(R)$ with respect to end-to-end delay
and \cite{OurUpperBoundPaper} tells us that we cannot do any better
than the sphere-packing bound without feedback. 
\begin{eqnarray}
E_r(R) & = & \max_{\rho \in [0,1]} E_0(C_f,\rho) - \rho
R C_f \label{eqn:randombound} \\
E_{sp}(R) & = & \max_{\rho \in [0,\infty)} E_0(C_f,\rho) - \rho R C_f
\label{eqn:spherebound}
\end{eqnarray}
where $R$ ranges from $0$ to the forward capacity of $1-\beta_f$
packets (of size $C_f$ each) per channel use. 

Using a fixed-length block code would introduce another factor of two
in end-to-end delay since the message to be transmitted would first
have to be buffered up to make a block.

\subsubsection{Pure feedback error correction}

The intuitive ``repeat until success'' strategy for perfect feedback
analyzed in \cite{OurUpperBoundPaper} can be adapted to when the
feedback is unreliable. For simplicity, focus on $k_f = k_b = 1$. The
idea is for the feedback encoder to use $1$ bit on the feedback
channel to indicate if the forward packet was received or not. If this
feedback packet is not erased, the situation is exactly as it is when
feedback is perfect. When the feedback packet is erased, the safe
choice for the forward encoder is to retransmit. However, this
requires some way for the decoder to know that the incoming packet is
a retransmission rather than a new packet. The practical way this
problem is solved is by having sequence numbers on packets. As
\cite{MasseyLecture} points out, only $1$ bit of overhead per forward
packet is required for the sequence number in this scenario.

Thus, the resulting system behaves like a repeat until success system
with perfect feedback with two modifications:
\begin{itemize}
 \item The effective forward packet size goes from $C_f$ bits to $C_f
   - 1$ bits to accommodate the $1$-bit sequence numbers.

 \item The effective erasure probability goes from $\beta_f$ to
   $(1-(1-\beta_f)(1-\beta_b))$ because an erasure on either forward or
   feedback channel demands a retransmission.
\end{itemize}
 
Thus an error exponent of $\alpha$ with respect to end-to-end delay
can be achieved as long as the rate $R$ (in units of $C_f$ bits at a
time) satisfies (using Theorem~3.3 in \cite{OurUpperBoundPaper} and
adjusting for $\alpha$ being in base $e$ rather than base $2$):
\begin{equation} \label{eqn:boundfornaivefeedbackstrategy}
R < (\frac{C_f - 1}{C_f} )
\frac{\alpha}{\alpha + \ln
  \left(\frac{(1-\beta_f)(1-\beta_b)}{1-\exp(\alpha)
 (1-(1-\beta_f)(1-\beta_b)) }\right)}
\end{equation}
and $0 < \alpha < -\ln (1-\beta_f)(1-\beta_b)$. No higher $\alpha$
can be achieved by this scheme. Even as $\alpha \rightarrow 0$ and
$C_f \rightarrow \infty$, the above rate only reaches
$(1-\beta_f)(1-\beta_b)$ and thus is bounded away from the capacity
$1-\beta_f$ of the forward link. 

\subsubsection{Comparison}

\begin{figure}
\begin{center}
\includegraphics[width=6in]{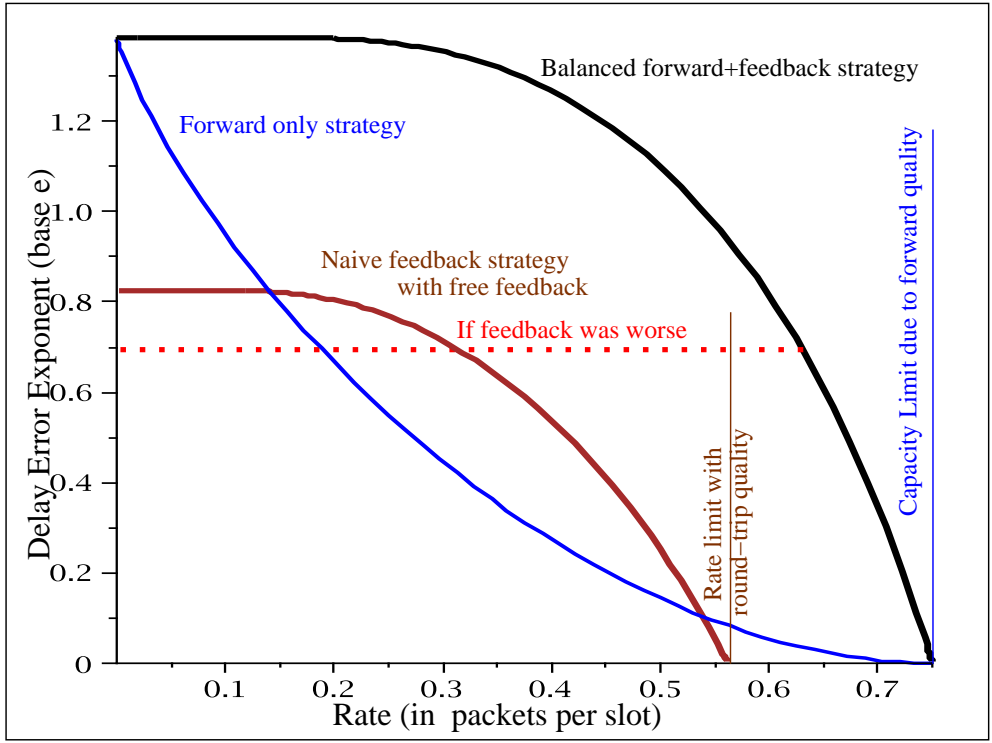}
\end{center}
\caption{The error exponents governing the asymptotic tradeoff between
  the probability of error and end-to-end delay for an $0.25$-erasure
  channel with a separate $0.25$-erasure channel on the feedback
  link. The top curve is the {\em uncertainty-focusing bound} from
  \cite{OurUpperBoundPaper} that is optimal assuming that feedback is
  perfect. If the probability of erasure on the reverse channel were
  increased to $0.50$, the achievable exponents for the schemes of
  this paper saturate at $\ln 0.5$. For comparison, the simple
  feedback-only (with $0.25$ erasure on both forward and feedback
  links) strategy achieves only the lower curve. The forward-only
  curve bounds what is possible without using feedback in general and
  also what is possible with feedback if the system is restricted to
  fixed-length block codes. }
\label{fig:delayrobust}
\end{figure}

Three scenarios are considered. Figure~\ref{fig:delayrobust} sets
$k_f=k_b=1$ and compares the pure feedback and pure forward error
correction strategies to the balanced approach of
Theorem~\ref{thm:nolistbasictheorem} when the erasure probability on
both the forward and feedback links are the same. The limit of $C_f
\rightarrow \infty$ is shown, although this is only significant at low
rates. If the feedback link were more unreliable than the forward
link, then the reliability gains from
Theorem~\ref{thm:nolistbasictheorem} would saturate at lower
rates. Looking at the curves in the vicinity of capacity shows clearly
that the factor reduction in asymptotically required end-to-end delay
over purely forward error correction tends to $\infty$.

\begin{figure}
\begin{center}
\includegraphics[width=6in]{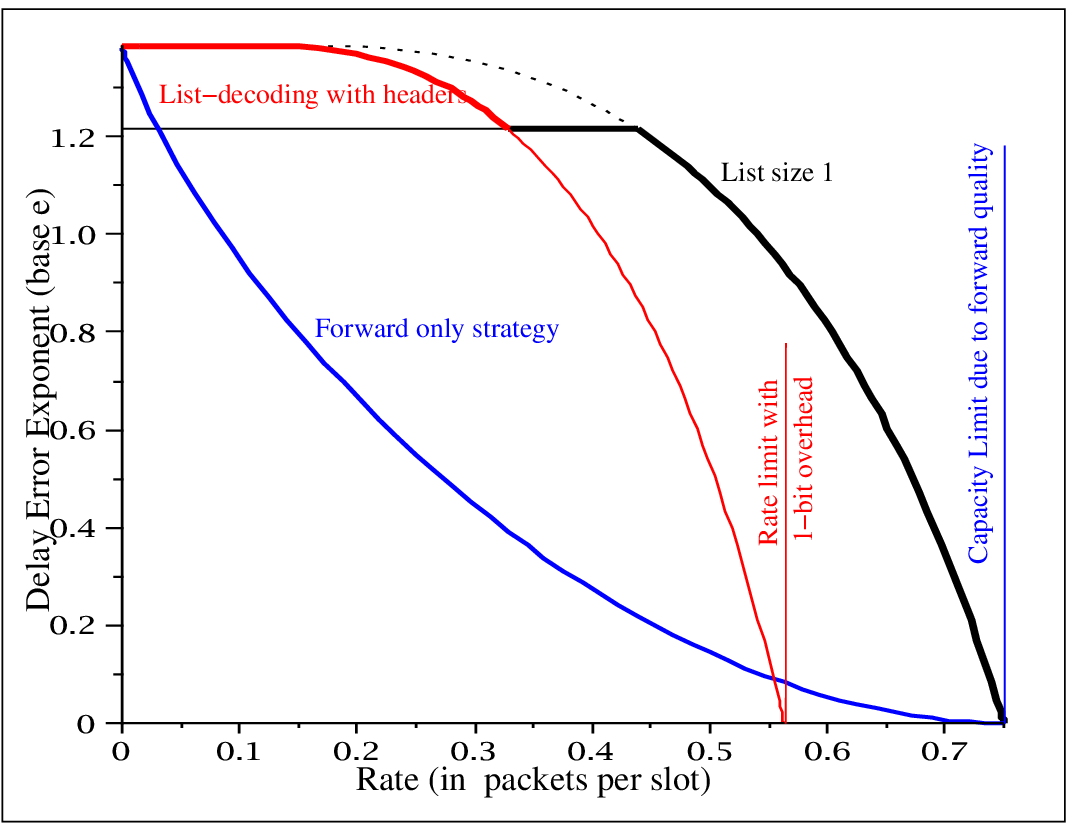}
\caption{The impact of list decoding is significant only at low
  rates. The top curve is the uncertainty-focusing bound. The dotted
  segment is the part that is not achieved. The red curve represents
  what is achievable using Theorem~\ref{thm:listbasictheorem} and the
  thick black curve represents
  Theorem~\ref{thm:nolistbasictheorem}. The discontinuity 
  in the thick curve indicates when it is worth switching between the
  two theorems. The forward packet size $C_f=4$ and the probability of
  erasure is $0.25$ on both the forward and feedback links. } \label{fig:listvsnolist} 
\end{center}
\end{figure}

Figure~\ref{fig:listvsnolist} illustrates the difference between
Theorems \ref{thm:nolistbasictheorem} and
\ref{thm:listbasictheorem}. For high rates,
Theorem~\ref{thm:nolistbasictheorem} is better. But at low rates,
Theorem~\ref{thm:listbasictheorem} provides better reliability and
hence shorter asymptotic end-to-end delays. When the packets are
short, the capacity penalty for allocating $1$ bit for a header can be
significant as the plot illustrates using $C_f=4$. 

\begin{figure}
\begin{center}
\includegraphics[width=6in]{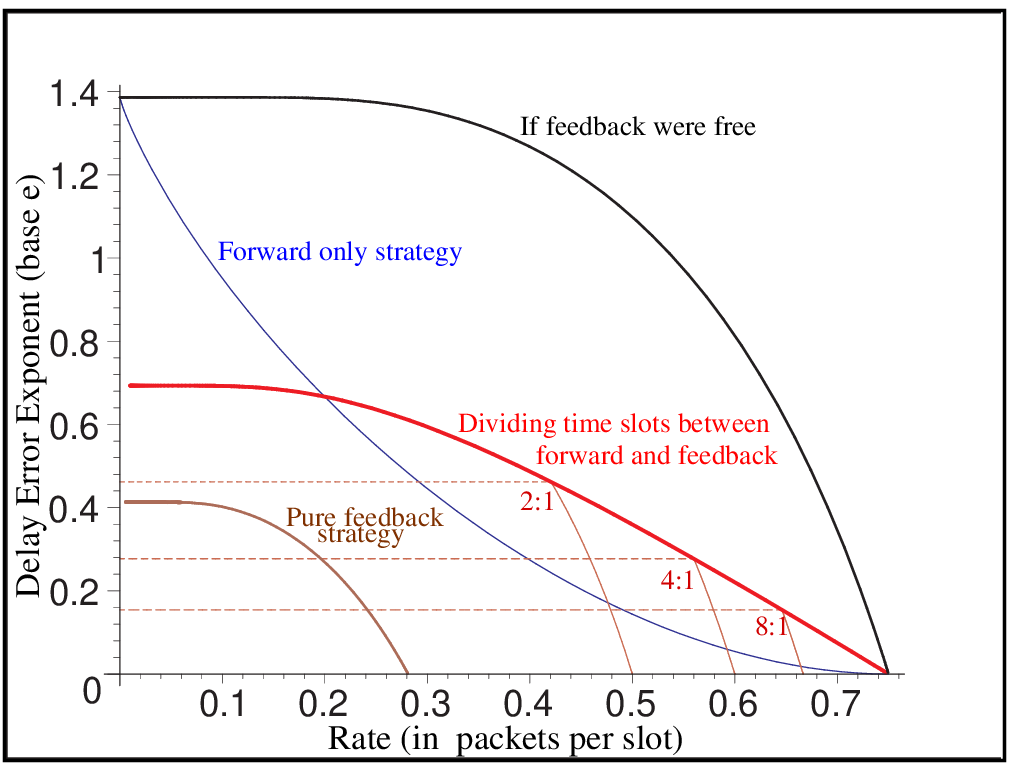}
\end{center}
\caption{The upper curve is the uncertainty-focusing bound with
  perfect and free feedback and the lower-most curve is the delay
  performance attained by feedback-only error-correction strategies
  that split the channel equally among forward and feedback uses. The
  intermediate curves reflect giving everything to the forward channel
  and various ratios of forward to feedback channel uses. The envelope
  of such schemes (described in Theorem~\ref{thm:balancedtheorem}) is
  also plotted.}
\label{fig:delayshared}
\end{figure}

Figure~\ref{fig:delayshared} illustrates the scenario of
Theorem~\ref{thm:balancedtheorem} in that it assumes that there is a
single shared physical channel that must be divided between forward
and feedback channel uses. Somewhat surprisingly, feedback becomes
more valuable the closer the system comes to capacity. The factor
reduction in asymptotically required end-to-end delay over purely
forward error correction tends to $\infty$ as the data rate approaches
capacity. This shows that, at least in the packet-erasure case,
feedback is worth implementing even if it comes at the cost of taking
resources away from the forward path.

\begin{figure}
\begin{center}
\includegraphics[width=6in]{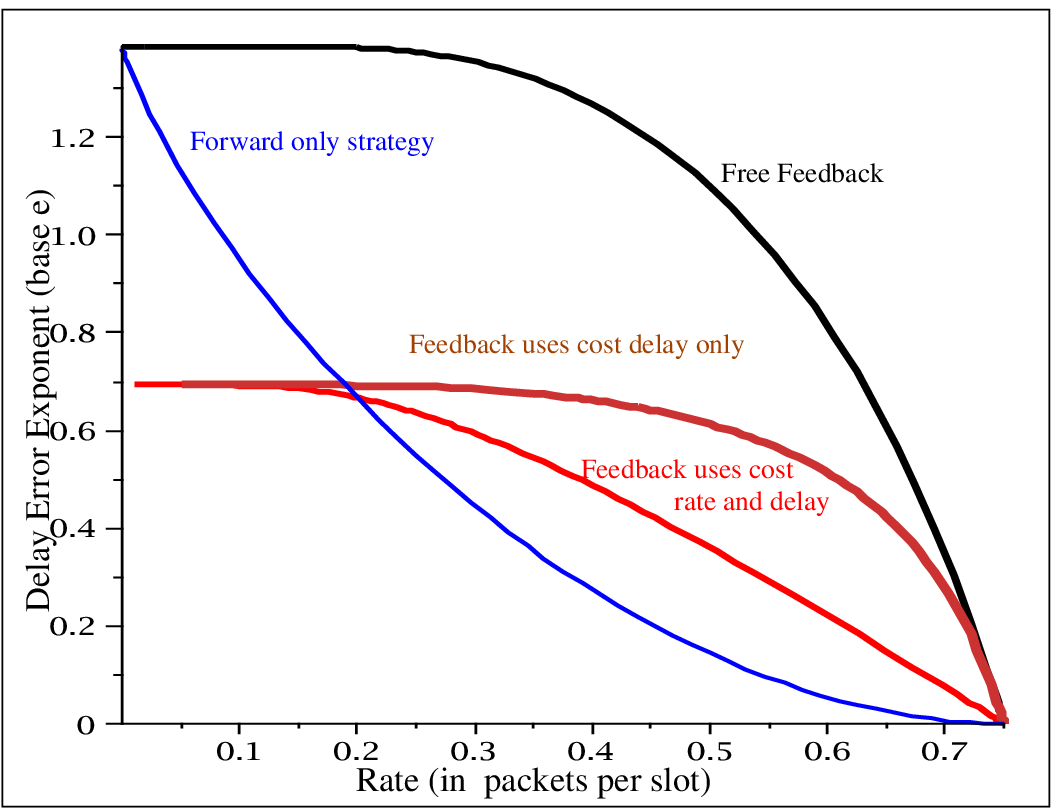}
\end{center}
\caption{The upper curve is the uncertainty-focusing bound with
  perfect and free feedback. The second curve down optimizes the split
  between forward and feedback channel uses and counts end-to-end
  delay in terms of total channel uses. Rate is calculated only
  relative to the forward channel uses with the idea that the feedback
  packets are carrying other useful data. The third curve is the one
  from Figure~\ref{fig:delayshared} that counts both delay and rate
  relative to total channel uses. The final curve is for forward
  error-correction only.} 
\label{fig:delaysharednoratepenalty}
\end{figure}

Finally, Figure~\ref{fig:delaysharednoratepenalty} illustrates the
impact of how rate and delay are counted. Notice that at high rates,
the curve in which the feedback is counted against the delay but not
the rate is very close to the case in which feedback is free. This
makes sense when the packets are large and there is presumably
independent data coming in the opposite direction.

\section{Unreliable, but ``free'' feedback}
\label{sec:freefeedback}

Throughout this section, the $k_f$ and $k_b$ are considered to be a
given. The goal is to prove Theorems \ref{thm:nolistbasictheorem} and
\ref{thm:listbasictheorem}. The basic idea is to adapt the $(n,c,l)$
scheme of \cite{OurUpperBoundPaper} to this situation. Corollary~6.1
of \cite{OurUpperBoundPaper} is the key tool used to prove the
results.

\subsection{Theorem~\ref{thm:nolistbasictheorem}: no list decoding}
\label{sec:nolistproof}

The scheme used is:
\begin{enumerate}
  \item Group incoming bits into blocks of size $nck_f R C_f$
        each. Assume that $n$ and $c$ are both large, but $nc k_f$ is
        small relative to the target end-to-end delay measured in
        forward channel uses.

  \item Hold blocks in a FIFO queue awaiting transmission. The first
        block is numbered $1$ with numbers incrementing by $1$
        thereafter. At time $0$, both sides agree that block $0$ has
        been decoded. The current pointer for both is set at $1$.

  \item The forward encoder transmits the oldest waiting block using
        an $\infty$-length random codebook (rateless code) with a new
        codebook being drawn for each block. The codewords themselves
        consist of iid uniform $C_f$-bit packets.

        Formally, the codewords are $X_{i}(j,t)$ where $i > 0$
        represents the current block number, $t > 0$ is the
        current time, and $0 \leq j < 2^{nc k_f R C_f}$ is the
        value of the current block.  Each $X_{i}(j,t)$ is drawn iid
        and is $k_f$ packets long. 

      \item The forward decoder uses the received channel symbols
        $Y(t)$ to eliminate potential messages (codewords) that could
        have been sent as the current block $i$. As soon as there is
        only one solitary codeword $j$ left, the decoder considers it
        to be the true value $\widehat{j}$ for that block and the
        block is marked as successfully decoded. When the current
        received packet $Y(t)$ is incompatible with this solitary
        codeword (ie $-1 \neq Y(t) \neq X_{i}(\widehat{j},t)$), then
        the current block count $i$ is incremented at the decoder and
        it considers the next block to have begun.
  
  \item The feedback encoder always uses its one bit to send back the
        modulo $2$ number of the last block (usually $i-1$, but
        sometimes $i$ when the current block has been decoded and the
        receiver is still waiting for a sign that the next block has
        begun) that was successfully decoded. 

  \item If the forward encoder receives feedback indicating that
        the current block has been successfully decoded, it removes
        the current block from the queue, increments the current $i$
        pointer, and moves on to the next block. If there are no
        blocks awaiting transmission, the encoder can just continue
        extending the current codeword until there is something new to
        send. 
\end{enumerate}

An interesting feature of this scheme is that {\em there are no
  explicit sequence numbers} on the forward packets unlike the
approach of \cite{MasseyLecture}. Instead, they are implicit. This
prevents a loss of rate. Synchronization between the forward encoder
side and decoder side is maintained because:
\begin{itemize}
 \item They start out synchronized.
 \item The forward encoder can only increment its pointer after getting explicit
       feedback from the decoder telling it that the block has been
       correctly received. Because the feedback channel is an erasure
       channel, this implies that such an acknowledgement was actually
       sent.
 \item The decoder can only increment its pointer $i$ after
       receiving a symbol that is incompatible with the prior
       codeword. Because the forward channel is an erasure channel,
       the only way this can happen is if the packet was indeed sent
       from a new codebook indicating unambiguously that the forward
       encoder has incremented its pointer.
\end{itemize}
It is thus clear that no errors are ever made. The total delay
experienced by a bit can be broken down into three parts:
\begin{enumerate}
 \item {\em Assembly delay}: How long it takes before the rest of the
  message block has arrived at the forward encoder. (Bounded by the
  constant $nck_f$ forward channel uses and hence asymptotically
  irrelevant as $d \rightarrow \infty$.)
 \item {\em Queuing delay}: How long the message block must wait
   before it begins to be transmitted. 
 \item {\em Transmission delay}: How many channel uses it takes
   before the codeword is correctly decoded (Random quantity $T'_i$
   that must be an integer multiple of $k_f$.) 
\end{enumerate}

To understand this, a closer look at the transmission delay $T'_i$ is
required. First, the $T'_i$ can be upper-bounded by the service time
$T_i$ that measures how long it takes till the forward encoder is sure that
the codeword was correctly decoded. This puts us in the setting of
Corollary~6.1 of \cite{OurUpperBoundPaper}. 

\subsubsection{The service time}
$T_i = T_{1,i} + T_{2,i} +  T_{3,i}$ consists of the sum of how long
it takes to complete three distinct stages of service.
\begin{itemize}
 \item $T_{1,i}$: How long till the decoder realizes that the forward
   encoder has moved on. 

   Since the new codeword's symbol $X_{i}(j,t)$ is drawn independently
   from the previous codeword's symbol $X_{i-1}(j_{prev},t)$, the
   probability of a received packet being ambiguous is just $\beta_f +
   (1-\beta_f)2^{-C_f}$ since there is a $\beta_f$ probability of
   being erased and only a $1$ in $2^{C_f}$ chance of 
   drawing something identical. Thus:
  \begin{eqnarray*}
  {\cal P}(T_{1,i} > t) 
  & = & (\beta_f + (1-\beta_f)2^{-C_f})^{t} \\
  & = & \exp(t \ln(\beta_f + (1-\beta_f)2^{-C_f})) \\
  & = & \exp(-t E_0(C_f,1)) 
  \end{eqnarray*}
  and so $T_{1,i}$ has a geometric distribution governed by the
  exponent $E_0(C_f,1)$. (Alternatively, this can be seen directly
  from the interpretation of $E_0(1)$ as the exponent governing the 
  pairwise probability of  confusion for codewords  for the regular
  union bound \cite{Gallager}.)
 
  The $T_{1,i}$ for different values of $i$ are clearly iid since the
  codebook is independently drawn at each time $t$ and the channel is
  memoryless. 

 \item $T_{2,i}$: How long until the decoder is able to decode the
   codeword uniquely.

  Lemma~7.1 of \cite{OurUpperBoundPaper} applies to this term without 
  list decoding. The $T_{2,i}$ are thus also iid across
  blocks $i$ and 
  \begin{equation} \label{eqn:boundforservice}
  {\cal P}(T_{2,i} - \lceil \widetilde{t}(\rho,R,n) \rceil ck_f > tck_f) 
\leq 
\exp\left(-tck_f E_0(C_f,\rho) \right)
 \end{equation}
 for all $\rho \in [0,1]$ and where
$\widetilde{t}(\rho,R,n) = \frac{R}{\widetilde{C}(\rho)}n$ and
$\widetilde{C}(\rho) = \frac{E_0(C_f,\rho)}{\rho C_f \ln 2}$ after
adjusting for the units of $C_f$ {\em bit} packets used here
for rate $R$. 

 \item $T_{3,i}$: How long until the encoder realizes that the decoder
   has moved on.

  The only way the encoder could miss this is if all the feedback
  packets were erased since the decoding succeeded. The only subtlety
  comes in measuring time. In keeping with tradition, in this section
  time is measured in forward channel uses and thus $k_f$ and $k_b$
  are needed to translate.
 \begin{eqnarray*}
  {\cal P}(T_{3,i} > t k_f) 
  & = & (\beta_b)^{t k_b} \\
  & = & \exp(-t k_b (-\ln \beta_b))
 \end{eqnarray*}
 and so the relevant exponent for $T_3$ is $- \frac{k_b}{k_f} \ln \beta_b$. 
\end{itemize}

$T_{3,i}$ depends on the randomness in the feedback channel while
$T_{1,i}$ and $T_{2,i}$ both depend on the forward channel and random
codebooks, but at disjoint times.  Thus, each of the terms are
independent of each other since they depend on different independent
random variables.

It is clear that the sum of three independent geometric random
variables can be bounded by the slowest of them plus a constant. If
two of them are equally slow, then the resulting polynomial growth
term can be absorbed into a slightly smaller exponent. Thus for all
$\epsilon > 0, \exists K$ depending on $\epsilon$ and the triple
$(-\frac{k_b}{k_f} \ln \beta_b, E_0(C_f,\rho), E_0(C_f,1))$ so that:
\begin{equation} \label{eqn:servicetimebound}
{\cal P}(T_{i} - \lceil \widetilde{t}(\rho,R,n) \rceil ck_f - K > tck_f) 
\leq 
\exp(-tck_f (\min(E_0(C_f,\rho), -\frac{k_b}{k_f} \ln \beta_b) - \epsilon))
\end{equation}
since $E_0(C_f,\rho) \leq E_0(C_f,1)$ because $\rho \in [0,1]$ and the
Gallager function $E_0$ is monotonically increasing in $\rho$. Notice
also that the constant $K$ here does not depend on $n$ or $c$.

\subsubsection{Finishing the proof}
The conditions of Corollary~6.1 of \cite{OurUpperBoundPaper} now apply
with point messages arriving every $nck_f$ channel uses, or every $n$
units of time if time is measured in increments of $ck_f$ channel
uses. At this point, the proof proceeds in a manner identical to that
of Theorem~3.4 in \cite{OurUpperBoundPaper}. As long as $\rho$ is
chosen small enough so that $R < \frac{E_0(C_f,\rho)}{\rho C_f \ln
  2}$, there exists $n$ large enough so that $\widetilde{t}(\rho,R,n)
+ \frac{K}{ck_f} < (1-\delta)n$. The effective rate $R''$ from
Corollary~6.1 of \cite{OurUpperBoundPaper} is thus $\frac{1}{n -
  \widetilde{t}(\rho,R,n) + \frac{K}{ck_f}} < \frac{1}{\delta n}$
point messages per $ck_f$ forward channel uses. This can be made
arbitrarily small by making $n$ large and so by Theorem~3.3 in
\cite{OurUpperBoundPaper} the error exponent with end-to-end delay can
be made arbitrarily close 
to $\min(E_0(C_f,\rho), -\frac{k_b}{k_f} \ln \beta_b)$. If the target
error exponent $\alpha$ satisfies (\ref{eqn:basicboundforalphalist1}),
then the minimum is known to be $E_0(C_f,\rho)$. This is maximized by
increasing $\rho$ so that $\frac{E_0(C_f,\rho)}{\rho C_f \ln 2}$
approaches $R$ from above. 

This demonstrates the asymptotic achievability of all
reliability/rate points within the region obtained by varying $\rho
\in [0,1]$: 
\begin{eqnarray}
\alpha &<& \min(-\frac{k_b}{k_f} \ln \beta_b, E_0(C_f, \rho)), \nonumber \\
R &<& \frac{E_0(C_f,\rho)}{\rho C_f \ln  2} \label{eqn:parametricfocusing}
\end{eqnarray}

Observe that $\rho C_f$ appears together in
the expression for $E_0$ in (\ref{eqn:Enoughtdefinition}) and so
$\rho C_f$ plays the role of simple $\rho$ for the binary erasure
channel. Theorem~3.3 in \cite{OurUpperBoundPaper} then gives the
desired expression for the performance after converting from base $2$
to base $e$.

The anytime property is inherited from Corollary~6.1 of
\cite{OurUpperBoundPaper}. \hfill $\Box$ \vspace{0.20in}

\subsection{Theorem~\ref{thm:listbasictheorem}: with list decoding}
\label{sec:listproof}

When the forward and reverse channels have packet sizes of at least
$2$, it is possible to augment the protocol to use list-decoding to a
list of size $\ell$ and some interaction to resolve the list
ambiguity. The idea is to have the feedback encoder tell the forward
encoder a subset of bit positions in the message block that it is
confused about. For any pair of distinct messages, there exists a
single bit position that would resolve the ambiguity between
them. Since there are $\ell$ messages on the list, $\frac{\ell (\ell -
  1)}{2}$ such bit positions are clearly sufficient. Once the forward
encoder knows which bit positions the decoder is uncertain about, it
can communicate those particular bit values reliably using a
repetition code.

The scheme is an extension of the scheme of
Theorem~\ref{thm:nolistbasictheorem} except with each message block requiring 
$m = 1 + \frac{\ell (\ell - 1)}{2}(1 + \lceil \log_2 n c k_f R C_f
\rceil)$ rounds of communication instead of just $1$ round. To support
these multiple rounds, $1$ bit is reserved on every forward packet and
every feedback packet to carry the round number modulo $2$. 

These rounds have the following roles:
\begin{itemize}
\item $1$ round: (Forward link leads, feedback link follows) A random
  codebook $X_{i}(j,t)$ as in the previous section is used by the
  forward encoder as before to communicate most of the information in
  the message block. The round stops when decoder has decoded to within $\ell$
  possible choices of the codeword $j$. At that point, the feedback encoder
  will increment the round count in the feedback packets and
  initiate the next round.

\item $\frac{\ell (\ell - 1)}{2} \lceil \log_2 n c k_f R C_f \rceil$
  rounds: (Feedback link leads, forward link follows) The feedback
  encoder uses a repetition code to communicate $\frac{\ell (\ell -
    1)}{2}$ different bit positions within the block. Since there are
  $n c k_f R C_f$ bits within a block, it takes $\lceil \log_2 n c k_f
  R C_f \rceil$ bits to specify a specific bit position. 

  The feedback encoder uses the second bit in each $2$-bit packet to
  carry the repetition code encoding these positions. As soon as the
  feedback channel is successful, the forward encoder will signal the
  round to advance by incrementing its counter. If this was the last
  such round, the forward encoder will initiate the next type of
  round. Otherwise, as soon as a forward channel is successful, the
  feedback encoder will also increment the round and move on to the
  next bit. 

\item $\frac{\ell (\ell - 1)}{2}$ rounds:  (Forward link leads,
  feedback link follows) The forward encoder uses a repetition
  code to communicate the specific values of the $\frac{\ell (\ell -
    1)}{2}$ requested bits. The rounds advance exactly as in the
  previous set of rounds.
\end{itemize}

Synchronization between the encoder and decoder is maintained because:
\begin{itemize}
 \item They start out synchronized.

 \item The follower advances its counter as soon as it has decoded the
 round. As soon as the leader hears this, it too moves on to the next
 round. Because each packet comes with an unmistakable counter, it is
 interpreted correctly.
\end{itemize}
 
Because the channels are erasure channels, there is no possibility for
confusion. Each follower advances to the next round only when it has
learned what it needs from this one. 

The proof that this achieves the desired error exponents is mostly
parallel to that of the previous section. In the interests of brevity,
only the differences are discussed here.

\subsubsection{The service time}
$T_i = T_{1,i} + \sum_{k=1}^{\frac{\ell (\ell - 1)}{2} \lceil \log_2 n
  c k_f R C_f \rceil} T_{2,i,k} + \sum_{k=1}^{\frac{\ell (\ell -
    1)}{2}} T_{3,i,k}$ since each round needs to complete for the
entire block to complete.  
\begin{itemize}

\item $T_{1,i}$: How long until the decoder is able to decode the
  codeword to within a list of size $\ell$. This is almost the same as
  $T_{2,i}$ in the previous section. The only difference is that the
  effective forward packet size is $C_f - 1$ bits since $1$ bit is
  reserved for the round number modulo $2$. 

  Lemma~7.1 of \cite{OurUpperBoundPaper} applies to this term with a
  list size of $\ell$. The $T_{1,i}$ are thus iid across blocks $i$ and 
  \begin{equation} \label{eqn:boundforservicelist}
  {\cal P}(T_{1,i} - \lceil \widetilde{t}(\rho,R,n) \rceil ck_f > tck_f) 
\leq 
\exp(-tck_f E_0(C_f -1,\rho) )
 \end{equation}
 for all $\rho \in [0,\ell]$ and where $\widetilde{t}(\rho,R,n) =
 \frac{R}{\widetilde{C}(\rho)}n$ and $\widetilde{C}(\rho) =
 \frac{E_0(C_f - 1,\rho)}{\rho C_f \ln 2}$ after adjusting for the
 notation used here including the units of $C_f$ {\em bit} packets for
 rate $R$. 

\item $T_{2,i,k}$: How long it takes to complete one round of
  communicating a single bit from the feedback encoder to the forward
  encoder. This is the sum of two independent geometric random
  variables: one counting how long till a successful use of the
  feedback channel carrying the bit, and a second counting how long
  till a successful use of the forward channel carrying the
  confirmation that the bit was received.

\item $T_{3,i,k}$: How long it takes to complete one round of
  communicating a single bit from the forward encoder. This is also
  the sum of the same two independent geometric random variables.
\end{itemize}

Use $T_f(k)$ to denote independent geometric (in increments of $k_f$)
random variables counting how long it takes till a successful use of
the forward channel. Similarly, use $T_b(k)$ for the backward channel
in increments of $k_b$. Thus, $T = T_1 + \sum_{k=1}^{m-1} (T_f(k) +
\frac{k_f}{k_b} T_b(k))$ has the distribution of the service time in
terms of forward channel uses.

Clearly $E_0(C_f -1,\rho) < -\ln \beta_f$ for all $\rho > 0$. There
are two possibilities depending on whether $T_1$ provides the dominant
error exponent: $E_0(C_f -1,\rho^*) < -\frac{k_b}{k_f} \ln \beta_b$ for
the $\rho^*$ that solves $\widetilde{C}(\rho^*) = R$. If the feedback
channel provides the dominant exponent, set $\rho < \rho^*$
so that $E_0(C_f -1,\rho) = -\frac{k_b}{k_f} \ln \beta_b$. Otherwise,
leave $\rho < \rho^*$ free for now. Define $\gamma = E_0(C_f -1,\rho)$
as the dominant exponent.

Let $T'(k)$ be iid geometric random variables (in increments of $c k_f$) 
that are governed by the exponent $\gamma$ so ${\cal P}(T' > t c k_f) =
\exp(-t c k_f \gamma)$. Consider $\sum_{k=1}^{2m-1}T'(k)$. This has a
negative binomial or Pascal distribution.

\begin{lemma} \label{lem:pascalbound}
Let $T'(k)$ be iid geometric random variables that are governed by the
exponent $\gamma$ so ${\cal P}(T' > t) = \exp(-t \gamma)$. Then, for
every $\epsilon' > 0$, there exists an $\epsilon > 0$ that depends
only on $\gamma$ and $\epsilon'$ so that $\forall t > 0$:
\begin{equation}
{\cal P}(\sum_{k=1}^{2m-1}T'(k) > t + \check{t}) < 
\exp(-\gamma(1-\epsilon') (t + \check{t}))
\end{equation}
where $\check{t} = \frac{2m - 1}{\epsilon}$. 
\end{lemma}
{\em Proof: }See Appendix~\ref{app:pascal}.
\vspace{0.1in}

This means that the service time $T_i$ has a complementary CDF that is 
bounded by:
\begin{equation} \label{eqn:boundforservicelisttotal}
  {\cal P}(T_{i} - (\lceil \widetilde{t}(\rho,R,n) \rceil  +
  \check{t}) ck_f > tck_f) 
<
\exp(-tck_f (1-\epsilon')E_0(C_f -1,\rho)).
\end{equation}

\subsubsection{Finishing the proof} The proof can continue almost the
same way as in the previous section. All that needs to be checked is
that for a given target error exponent $(1-\epsilon')E_0(C_f - 1,
\rho)$, the overhead $\lceil \widetilde{t}(\rho,R,n) \rceil  +
  \check{t}$ can be made smaller than $n$ so that the point message
  rate $R''$ in Corollary~6.1 of \cite{OurUpperBoundPaper} can be made
  to go to zero. 

Assuming that $\ell \geq 2$ (otherwise what is the point of using
list-decoding!):
\begin{eqnarray*}
\lceil \widetilde{t}(\rho,R,n) \rceil  + \check{t} 
& \leq & 1 + \frac{R}{\widetilde{C}(\rho)}n + \frac{2 + 2\frac{\ell (\ell - 1)}{2}(1 + \lceil \log_2 n c k_f R C_f
\rceil)}{\epsilon} \\
& \leq & [\frac{R}{\widetilde{C}(\rho)} + \frac{\ell (\ell -
  1)(3 + \log_2 c k_f R C_f )}{\epsilon} (\frac{\log_2 n}{n})]n.
\end{eqnarray*}
Clearly whenever $\widetilde{C}(\rho) > R$, there exists an $n$ big
enough so that the entire term in brackets $[\cdots] < 1-\delta$ for
some small $\delta > 0$. From this point on, the proof proceeds
exactly as before. Recall that $\epsilon'$ is arbitrary so this gets
us asymptotically to the fixed-delay reliability region parametrized as:
\begin{eqnarray}
 \alpha & < & \min(E_0(C_f - 1, \rho), -\frac{k_b}{k_f} \ln \beta_b), \nonumber \\
 R & < & \frac{E_0(C_f - 1, \rho)}{\rho C_f \ln 2} \label{eqn:listregion}
\end{eqnarray}
where $\rho$ ranges from $0$ to $\ell$. But $\ell$ can be chosen as high
as needed. Finally, the rate in (\ref{eqn:listregion}) can be
rewritten as $R < \frac{C_f - 1}{C_f} (\frac{E_0(C_f - 1, \rho)}{\rho
  (C_f -1) \ln 2})$. Notice that $(C_f - 1)\rho$ appear together in
the expression (\ref{eqn:Enoughtdefinition}) for $E_0(C_f - 1, \rho)$ 
in the place of the simple $\rho$ for the binary erasure channel. This
lets us use Theorem~3.3 in \cite{OurUpperBoundPaper} to get the
desired expression, once again doing the straightforward conversions
from base $2$ to base $e$. \hfill $\Box$ \vspace{0.20in} 

\subsection{Extensions}

While packet-erasure channels were considered for concreteness of
exposition, it is clear that Theorem~\ref{thm:nolistbasictheorem}
extends to any channel on the forward link for which the
zero-undetected-error\footnote{Zero-undetected-error means that the
  probability of error is zero if the decoder is also allowed to
  refuse to decode. For capacity to be meaningful, the probability of
  such refusals must approach zero as the delay or block-length gets
  large.} capacity equals the regular capacity (See Problem 5.32 in
\cite{csiszarkorner}). If the probability of undetected error is zero,
then decoding proceeds by eliminating codewords as being
impossible. That is all that is needed in this proof. In particular,
the result extends immediately to packet-valued channels that can
erase individual bits within packets according to some joint
distribution rather than having to erase only the entire packet or
nothing at all.

Similarly on the feedback path, the proof of
Theorem~\ref{thm:nolistbasictheorem} only requires the ability to
carry a single bit message unambiguously in a random amount of time,
where that time has a distribution that is bounded by a geometric. For
a channel whose zero-undetected-error capacity equals the regular
capacity, a random code can be used but with only two codewords. This
gives an exponent of at least $E_{0}^b(1)$ on the feedback
path. Therefore, the arguments of this section have essentially
already proved the following theorem showing optimality at high rates
for more general channels:

\begin{theorem} \label{thm:nolistgeneraltheorem} Consider the
  $(k_f,k_b)$ problem of Figures \ref{fig:feedbackblockdiagram} and
  \ref{fig:feedbacktimeline} with $k_f,k_b \geq 1$, and forward DMC
  $P_f$ and backward DMC $P_b$, both with their zero-undetected-error
  capacities (without feedback\footnote{Perfect feedback increases the
    zero-undetected-error capacity all the way to the Shannon capacity
    for DMCs. A system can just use a one bit message at the end to
    tell the decoder whether or not to accept its tentative decoding
    \cite{ForneyErasure}.}) equal to 
  their regular Shannon capacities. Suppose that $k_r > 0$ is the
  round-trip delay (measured in cycles). 

In the limit of large end-to-end delays, it is possible to
asymptotically achieve all $(R,\alpha)$ pairs 
\begin{eqnarray} 
\alpha &<& \min\left(-\frac{k_b}{k_f} E_0^b(1), E_0^f(\rho)\right), \nonumber \\
R &<& \frac{E_0^f(\rho)}{\rho} \label{eqn:generalboundlist1}
\end{eqnarray}
in an anytime fashion where $E_0^f$ and $E_0^b$ are the Gallager
functions for the forward and reverse channels respectively and $\rho
\in [0,1]$. The rate $R$ above is measured in nats per forward channel
use as is the reliability $\alpha$.
\end{theorem}
{\em Proof: }To deal with the round-trip delay, just extend the cycle
length by considering $k_f' = \frac{1}{\epsilon} k_f k_r, k_b' =
\frac{1}{\epsilon} k_b k_r$. Consider the last $k_f k_r$ channel uses
of each extended cycle to be wasted. The effective number of forward
channel uses is thus reduced by a factor $(1-\epsilon)$. Since this
factor reduction can be made as small we like, asymptotically, the
problem reduces to the case with no round-trip-delay. 

To patch the proof of Theorem~\ref{thm:nolistbasictheorem} to account
for the general channels on the forward and feedback links:
\begin{itemize}
 \item Use the $E_0^f(\rho^*)$ optimizing input distribution
   $\vec{q}(\rho^*)$ for the random forward channel codebooks. $\rho^*
   \in [0,1]$ is chosen so that the target $\alpha < E_0^f(\rho^*)$
   and $R < \frac{E_0^f(\rho^*)}{\rho^*}$. 

 \item Use the $E_0^b(1)$ optimizing input distribution for the
   random two-codeword codebooks on the feedback channel. 

 \item The analysis of $T_{1,i}$ is unchanged since
   $E_0^f(1,\vec{q}(\rho^*)) > E_0^f(\rho^*)$ by the properties of the
   Gallager function \cite{Gallager}. 

 \item The analysis of $T_{2,i}$ is entirely unchanged and gives
   $E_0^f(\rho^*)$ as the relevant exponent.

 \item The analysis of $T_{3,i}$ is now exactly parallel to $T_{1,i}$
   since this also succeeds the instant the two random codewords on
   the feedback link can be distinguished by the received symbol. This
   is governed by the exponent $E_0^b(1)$ in terms of feedback channel
   uses and thus $\frac{k_b}{k_f} E_0^b(1)$ in terms of forward
   channel uses.
 \end{itemize} 
 Everything else proceeds identically, except with rate units in nats
 per forward channel use rather than in normalized units of $C_f$
 bits. \hfill $\Box$ \vspace{0.20in}

 It is unclear how to extend Theorem~\ref{thm:listbasictheorem} to
 these general erasure-style channels. To break the $E_0(1)$ barrier,
 the construction in Theorem~\ref{thm:listbasictheorem} relies on
 having a single header bit that always shows up. This approach does
 extend to the packet-truncation channels of \cite{Nuno} by making the
 header bit come first, but it does not extend to erasure-style
 channels in which individual bits within a packet can be erased in
 some arbitrary fashion.

 The restriction to channels whose zero-undetected-error capacity
 without feedback {\em equals} their Shannon capacity is quite strict,
 and is required for the above schemes to work. This allows us to use
 imperfect feedback since the decoder can be counted on to know when
 to stop on its own. However, the approach of
 Section~\ref{sec:listproof} can be used to extend Theorem~3.4 of
 \cite{OurUpperBoundPaper} when the feedback is perfect. Recall that
 Theorem~3.4 of \cite{OurUpperBoundPaper} requires a zero-error
 capacity that is strictly greater than zero. The multi-round approach
 with repetition codes allows us to drop this condition and merely
 require that the zero-undetected-error capacity without feedback be
 strictly greater than zero ({\em ie} the channel matrix $P$ has at
 least one zero entry in a row and column that is not identically
 zero).
 
\begin{theorem} 
  For any DMC whose transition matrix $P$ contains a nontrivial zero,
  it is possible to use noiseless perfect feedback and randomized
  encoders to asymptotically approach all delay exponents within
  the region 
 \begin{eqnarray} \label{eqn:symmetricfeedbackbound}
 \alpha & < & \min\left(E_0(\rho), E^* \right),\\
 R & < & \frac{E_0(\rho)}{\rho} \nonumber
 \end{eqnarray}
  where $E_0(\rho)$ is the Gallager function, $\rho$ ranges from
  $0$ to $\infty$, and $E^*$ is the error exponent governing the
  zero-undetected-error transmission of a single bit message.

  Furthermore, the delay exponents $\alpha$ can be achieved in a
  delay-universal or ``anytime'' sense even if the feedback is delayed
  by an amount $\phi$ that is small relative to the asymptotically
  large target end-to-end delay.
\end{theorem}
{\em Proof: } The proof and overall approach is almost identical to
what has done previously. Only the relevant differences will be
covered here.

First, it is well known that the presence of a nontrivial zero makes
the zero-undetected-error capacity strictly positive. To review, let
$x,x'$ be input letters so that there exists a $y$ such ${\cal P}(Y =
y| X=x) = 0$ while ${\cal P}(Y = y| X=x') > 0$. Then the following two
mini-codewords can be used: $(x,x')$ and $(x',x)$. The probability of
unambiguous decoding is at least ${\cal P}(Y = y| X=x')$ and so by
using this as a repetition code, the error exponent $E^* \geq
\frac{-1}{2} \ln(1-{\cal P}(Y = y| X=x')) > 0$ per forward channel
use. Of course, it can be much higher depending on the specific
channel. 

The $(n,c,l)$ scheme of \cite{OurUpperBoundPaper} is modified as
follows: 
\begin{itemize}
\item Each chunk of $c$ channel uses is now itself implemented as
  variable length. It consists of a fixed $c_f$ channel uses
  that are used exactly as before to carry a part of a random
  codeword. To this is appended a variable-length code that
  uses perfect feedback to communicate exactly $1$ bit without error.
  This bit consists of the ``punctuation'' telling the decoder 
  whether or not there is a comma after this chunk ({\em ie} whether
  the decoder's tentative list-decoding contains the true codeword or
  not). 

 \item If the list size $\ell = 2^l \neq 1$, then the $l$ bits of
   list-disambiguation information are conveyed by using $l$
   successive single-bit variable-length codes before the next block
   begins. 
\end{itemize}

Perfect noiseless feedback is assumed as in \cite{OurUpperBoundPaper}
so that the forward encoder knows when to stop each round and move on.
No headers are required. The main difference from
Theorem~\ref{thm:listbasictheorem} is that the number of rounds
required to communicate a message block is not fixed. Instead, each
message takes a variable number of rounds. 

The idea is to choose $c - c_f$ large and then to pick an effective
$c_f$ that is so big that it is almost proportional to $c$. Let $T_i$
be the total service time for the $i$-th message block as measured in
forward channel uses. It is clear that this is the sum of
\begin{itemize}
 \item $T_{1,i} c_f$: The number of forward channel uses required
   for the random codeword before the message can be correctly
   list-decoded to within $\ell$ possibilities. This is exactly as
   before and is governed by Lemma~7.1 of \cite{OurUpperBoundPaper}.

 \item $\sum_{k=1}^{T_{1,i}}T_{2,i,k}$: The number of forward channel
   uses required to communicate the $T_{1,i}$ distinct punctuation
   symbols in the block.  Each individually is governed by the
   exponent $E^*$.

   This is the qualitatively new term.
 
 \item $\sum_{k=1}^{l}T_{3,i,k}$: The number of forward channel
   uses required to communicate all of the $l$ distinct $1$-bit
   disambiguation messages.  These are also governed by the exponent
   $E^*$. 
\end{itemize}

It is easiest to upper-bound the complementary CDF for $T_{1,i}c_f +
\sum_{k=1}^{T_{1,i}}T_{2,i,k} + \sum_{k=1}^{l}T_{3,i,k}$
together. Define $\epsilon = \frac{1}{c-c_f}$,
$\widetilde{t}(\rho,R,n) = \lceil \frac{R}{\widetilde{C}(\rho)}n
\rceil$. 
\begin{eqnarray*}
& & {\cal P}(T_{1,i}c_f + \sum_{k=1}^{T_{1,i}}T_{2,i,k} 
+ \sum_{k=1}^{l}T_{3,i,k}
-
\widetilde{t}(\rho,R,n) c > tc) \\
& \leq_{a} &
{\cal P}(T_{1,i}c_f - \widetilde{t}(\rho,R,n) c_f \geq t c_f)  \\
& &
+
\sum_{s=1}^{t} \left(
{\cal P}(T_{1,i}c_f - \widetilde{t}(\rho,R,n) c_f = (t-s)c_f) \right)
\left(
{\cal P}(\sum_{k=1}^{\widetilde{t}(\rho,R,n) + t-s + l}T_{2,i,k} > \widetilde{t}(\rho,R,n)(c-c_f) +
(t-s) (c-c_f) + sc) \right) \\
& \leq_{b} &
\exp(-tc_f E_0(\rho)) +
\sum_{s=1}^{t} \exp(-(t-s)c_f E_0(\rho)) 
\left( 
{\cal P}(\sum_{k=1}^{\widetilde{t}(\rho,R,n) + t-s + l}T_{2,i,k} >
(\widetilde{t}(\rho,R,n) + t - s) (c-c_f) + sc) \right) \\
& = & 
\exp(-tc_f E_0(\rho)) +
\sum_{s=1}^{t} \exp(-(t-s)c_f E_0(\rho)) 
\left( 
{\cal P}(\sum_{k=1}^{\widetilde{t}(\rho,R,n) + t-s + l} T_{2,i,k} >
\frac{\widetilde{t}(\rho,R,n) + t - s + l}{\epsilon} + sc - \frac{l}{\epsilon}))
\right) \\
& <_{c} & 
\exp(-tc_f E_0(\rho)) +
\sum_{s=1}^{t} \exp(-(t-s)c_f E_0(\rho)) 
\left( \exp(-(1-\epsilon')(\frac{\widetilde{t}(\rho,R,n) + t - s +
  l}{\epsilon} + sc - \frac{l}{\epsilon})E^*)
\right) \\
& =_{d} & \exp(-tc_f E_0(\rho)) +
\left( 
\exp(-(1-\epsilon')\frac{\widetilde{t}(\rho,R,n) + t}{\epsilon}E^*)
\right)
\sum_{s=1}^{t} \exp(-(t-s)c_f E_0(\rho)) 
\exp(-(1-\epsilon') s c_f E^*) \\
& <_{e} & 2\exp\left(-tc_f \min(E_0(\rho),(1-\epsilon')E^*) \right) \\
& =_{f} & 2\exp\left(-tc (\frac{c_f}{c}) \min(E_0(\rho),(1-\epsilon')E^*) \right).
\end{eqnarray*}
where $(a)$ is a union bound over different ways that the budget of
$\widetilde{t}(\rho,R,n) c + tc$ channel uses could be exceeded
together with the independence of the different component service
times. Notice that all the terms governed by $E^*$ are folded in
together. $(b)$ comes from simple algebra together with applying
Lemma~7.1 of \cite{OurUpperBoundPaper} and is valid as long as $\rho
\leq \ell$. $(c)$ is the result of substituting in the definition of
$\epsilon$ and then applying Lemma~\ref{lem:pascalbound}. $(d)$ brings
out the $sc_f$ term in the second term and then $(e)$ reflects that
the sum of exponentials is dominated by the largest term. $(f)$ is a
simple renormalization to $c$ units so that the result is plug-in
compatible with Lemma~7.1 of \cite{OurUpperBoundPaper}.

Choosing $c_f$ large enough tells us that for all $\epsilon'' > 0$,
there exist $c,c_f$ large enough so that:
\begin{equation} \label{eqn:newservicetimebound}
{\cal P}(T_j - \widetilde{t}(\rho,R,n) c > t c) \leq 
2[\exp\left(-c (1-\epsilon'')\min(E_0(\rho),E^*) \right)]^t
\end{equation}
as long as $\rho \in [0,\ell]$. From this point onward, the proof is
identical to the original in \cite{OurUpperBoundPaper}. \hfill $\Box$ \vspace{0.1in}

\section{Splitting a shared resource between the forward and feedback
  channel: Theorem~\ref{thm:balancedtheorem}}
\label{sec:costlyfeedback} 

The goal of this section is to prove Theorem~\ref{thm:balancedtheorem}
by considering what the best choices for $k_f$ and $k_b$ are if both
the rate and delay are considered relative to the sum $k_f+k_b$ rather
than just forward channel uses alone. 

\subsection{Evaluating the previous schemes}
Assume $k_f$ and $k_b$ are fixed and let $\eta_f = \frac{k_f}{k_f +
  k_b}$ and $\eta_b = \frac{k_b}{k_f + k_b}$. The $\eta_f$ acts as the
conversion factor mapping both the error exponents and the rates from
per-forward-channel-use units to per-total-channel-uses
units. Similarly, let $\xi_f = \frac{k_f C_f}{k_f C_f + k_b C_b}$ and
$\xi_b = \frac{k_b C_b}{k_f C_f + k_b C_b}$. The $\xi_f$ is the
conversion factor that maps error exponents and rates to
per-weighted-total-channel-uses units.

Thus for Theorem~\ref{thm:nolistbasictheorem},
(\ref{eqn:parametricfocusing}) becomes
\begin{eqnarray} 
\alpha' & < & \min\left(-\eta_b \ln \beta_b, \eta_f E_0(C_f, \rho)\right), \label{eqn:basicboundforalphaprimelist1} \\
\bar{\alpha} & < & \min\left(-\frac{k_b C_f}{k_b C_b + k_f C_f} \ln \beta_b, \xi_f E_0(C_f,\rho)\right), \label{eqn:basicboundforalphadoubleprimelist1} \\
R' &<& \frac{\eta_f E_0(C_f,\rho)}{\rho C_f \ln 2}, \label{eqn:parametricfocusingprime} \\
\bar{R} &<& \frac{\xi_f E_0(C_f,\rho)}{\rho C_f \ln 2}. \label{eqn:parametricfocusingdoubleprime}
\end{eqnarray}

For Theorem~\ref{thm:listbasictheorem}, the range of $\rho$
expands to $\rho \in [0,\infty)$ but the rate terms change to
\begin{eqnarray} 
R' &<& \frac{\eta_f E_0(C_f - 1,\rho)}{\rho C_f \ln 2}, \label{eqn:parametricfocusingprimelist} \\
\bar{R} &<& \frac{\xi_f E_0(C_f - 1,\rho)}{\rho C_f \ln 2}. \label{eqn:parametricfocusingdoubleprimelist}
\end{eqnarray}

\subsection{Optimizing by adjusting $k_f$ and $k_b$}

If $-\eta_b \ln \beta_b > \eta_f E_0(C_f, \rho))$, then
it is the forward link that is the bottleneck. If the inequality is in
the opposite direction, then the feedback link is what is
limiting reliability. This suggests that setting the two exponents equal to each
other gives a good exponent $\alpha'$. Since $\eta_b = 1-\eta_f$, this
means $\eta_f^* = \frac{-\ln \beta_b}{E_0(C_f,\rho) - \ln
  \beta_b}$. Plugging this in reveals that all $\alpha'$ are
achievable that satisfy:
\begin{eqnarray}
\alpha' & < & \frac{(-\ln \beta_b)E_0(C_f,\rho)}{E_0(C_f,\rho) - \ln
  \beta_b} \\
 & = & (\frac{E_0(C_f,\rho) - \ln \beta_b}{(-\ln
   \beta_b)E_0(C_f,\rho)})^{-1} \\
 & = & E_0'(C_f,\rho).
\end{eqnarray}
Plugging in for $R'$ reveals that all the $R'$ that satisfy
\begin{eqnarray}
R' & < & \frac{\eta_f E_0(C_f,\rho)}{\rho C_f \ln 2} \\
  & = & \frac{E_0'(C_f,\rho)}{\rho C_f \ln 2}
\end{eqnarray}
are also achievable. This establishes (\ref{eqn:balancedRegionNoList})
and identical arguments give (\ref{eqn:balancedRegionList}). 

To see what happens when $C_f$ gets large while $C_b$ stays constant,
just notice that in such a case $\xi_f = \frac{k_f C_f}{k_f C_f + k_b
  C_b}$ gets close to $1$ no matter how big $k_b$ is. This establishes
the $\alpha',\bar{R}$ tradeoff in (\ref{eqn:balancedRegionMixedTradeoff}).

Alternatively, $k_b$ can be chosen to be large enough so that
$-\frac{k_b C_f}{k_b C_b + k_f C_f} \ln \beta_b > - \xi \ln \beta_f >
\xi_f E_0(C_f, \rho)$. Then taking $C_f \rightarrow \infty$
immediately gives the desired $\bar{\alpha},\bar{R}$ tradeoff. \hfill $\Box$
\vspace{0.20in}

Theorem~\ref{thm:balancedtheorem} can clearly be extended to the
general setting of Theorem~\ref{thm:nolistgeneraltheorem}. The
relevant $E_0'(\rho)$ is immediately seen to be
\begin{equation} \label{eqn:generalBalancedEnought}
E_0'(\rho) = \left(((E_0^b(1))^{-1} + (E_0^f(\rho))^{-1} \right)^{-1}.
\end{equation}
The parallel to Theorem~3.5 of \cite{OurUpperBoundPaper} is obvious
and makes sense since both involve splitting a shared resource to two
purposes that must be balanced. Here, it is channel uses across the
feedback and forward channels. In Theorem~3.5 of
\cite{OurUpperBoundPaper}, it is allocating forward channel uses to
carrying messages and flow control information.

\section{Conclusions}

It has been shown that in the limit of large end-to-end delays,
perfect feedback performance is attainable by using appropriate random
codes at high rate for erasure channels even if the feedback channel
is an unreliable erasure channel. Somewhat surprisingly, this does not
require any explicit header bits on the packets if the rate is high
enough and thus works even for a system with a BEC in the forward link
and a BEC in the feedback link. The reliability gains from using
feedback are so large that they persist even when each feedback
channel use comes at the cost of not being able to use the forward
channel (half-duplex). This was shown by considering both rate and
delay in terms of total channel uses rather than just the forward
channel uses.

The arguments here readily generalize to all channels for which the
zero-undetected-error capacity equals the regular capacity, but do not
extend to channels like the BSC. Even when the zero-undetected-error
capacity is strictly larger than zero, the techniques here just give
an improved result for the case of perfect feedback. Showing that the
gains from feedback are robust to unreliable feedback in such cases
remains an open problem. In addition, the results here are on the
achievability side. The best upper-bounds to reliability in the fixed
end-to-end delay context are still those from
\cite{OurUpperBoundPaper} and it remains an open problem to tighten
the bounds when the feedback is unreliable or in the half-duplex
situation.

\appendices

\section{Pascal distribution bound: Lemma~\ref{lem:pascalbound}} \label{app:pascal} 

Consider $\sum_{k=1}^{2m-1}T'(k)$ where the $T'(k)$ are iid geometric random variables
with exponent $\gamma$. This has a negative binomial or Pascal
distribution. The probability distribution of this sum is easily
bounded by interpreting the Pascal distribution as the $(2m-1)-th$
arrival time of a virtual $\{Z_k\}$ Bernoulli process with probability
of failure $\exp(-\gamma c k_f)$. Pick an $\epsilon > 0$.

It is clear that
\begin{eqnarray*}
{\cal P}(\sum_{k=1}^{2m-1}T'(k) - \check{t} > t )
& = & 
{\cal P}(\sum_{k=1}^{t + \check{t}} Z_k  < 2m - 1) \\
& = & 
{\cal P}(\frac{\sum_{k=1}^{t + \check{t}} Z_k}{t + \check{t}}  <
\frac{2m - 1}{t + \check{t}}) \\
& \leq & 
{\cal P}(\frac{\sum_{k=1}^{t + \check{t}} Z_k}{t + \check{t}}  <
\frac{2m - 1}{\check{t}}) \\
& = & 
{\cal P}(\frac{\sum_{k=1}^{t + \check{t}} Z_k}{t + \check{t}}  <
\epsilon) \\
& \leq & 
(t + \check{t})\exp\left(-(t+\check{t})D\left(1-\epsilon||\exp(-\gamma)\right)\right).
\end{eqnarray*}
But when $\epsilon$ is small, 
\begin{eqnarray*}
D(1-\epsilon||\exp(-\gamma))
& = & 
(1-\epsilon)(\ln(1-\epsilon) + \gamma )
+ \epsilon(\ln(\frac{1}{1-\exp(-\gamma)}) + \ln \epsilon ) \\
& \geq &
(1-\epsilon)(\gamma - 2 \epsilon) + \epsilon \ln \epsilon  \\
& = & 
(1-\epsilon)\gamma - \epsilon(\ln\frac{1}{\epsilon} + 2(1-\epsilon)).
\end{eqnarray*}
So:
\begin{eqnarray*}
{\cal P}(\sum_{k=1}^{2m-1}T'(k) - \check{t} > t )
& \leq & 
(t + \check{t})\exp(-(t+\check{t})D(1-\epsilon||\exp(-\gamma))) \\
& \leq & 
\exp(-(t+\check{t})[(1-\epsilon)\gamma -
\epsilon(\ln\frac{1}{\epsilon} + 2(1-\epsilon)) - \frac{\ln
  (t+\check{t})}{t + \check{t}}]) \\
& < & 
\exp(-(t+\check{t})[(1-\epsilon)\gamma -
\epsilon(\ln\frac{1}{\epsilon} + 2) - \frac{\ln
  \check{t}}{\check{t}}]) \\ 
& = & 
\exp(-(t+\check{t})[(1-\epsilon)\gamma -
\epsilon(\ln\frac{1}{\epsilon} + 2 + \frac{\ln(2m-2) +
  \ln\frac{1}{\epsilon}}{2m-2})]) \\
& < & 
\exp(-(t+\check{t})[(1-\epsilon)\gamma -
\epsilon(2 \ln\frac{1}{\epsilon} + 3)])
\end{eqnarray*}
As $\epsilon \rightarrow 0$, the term $\epsilon(2\ln\frac{1}{\epsilon}
+ 3)$ also vanishes. So for any $\epsilon' > 0$, we can choose
$\epsilon$ small enough so that $(1-\epsilon)\gamma  -
\epsilon(2 \ln\frac{1}{\epsilon} + 3) > (1-\epsilon')\gamma$. This
gives:
\begin{eqnarray*}
{\cal P}(\sum_{k=1}^{2m-1}T'(k) - \check{t} > t  )
& < & 
\exp(-(t+\check{t})[(1-\epsilon)\gamma  -
\epsilon(2 \ln\frac{1}{\epsilon} + 3)]) \\
& < & 
\exp(-(t+\check{t})(1-\epsilon')\gamma)
\end{eqnarray*}
This completes the proof of Lemma~\ref{lem:pascalbound}. \hfill $\Box$

\section*{Acknowledgments}
We thank the NSF ITR program (ANI-0326503) for funding this
research. 

\bibliographystyle{IEEEtran}
\bibliography{IEEEabrv,./MyMainBibliography} \end{document}